\documentclass{elsart1p}

\usepackage{amssymb}
\usepackage{graphicx}

\renewcommand{\d}{\partial}
\newcommand{\eqref}[1]{(\ref{#1})}
\newcommand{\nn}{\nonumber\\}
\newcommand{\rh}{\varrho}
\newcommand{\ph}{\varphi}
\newcommand{\pint}[2]{{\int\!\frac{d^{#1}#2}{(2\pi)^#1}\,}}

\begin{document}

\begin{frontmatter}



\title{Renormalization and resummation in the O(N) model}


\author{A. Jakovac\thanksref{a}}
\thanks[a]{e-mail:jakovac\@esr.phy.bme.hu}
\address{Department of Physics, University of Wuppertal, D-42097 Wuppertal,
  Germany\thanksref{b}} 
\thanks[b]{On leave from Institute of Physics, BME Technical
  University, H-1521 Budapest, Hungary}

\begin{abstract}
  In the O(N) model for the large N expansion one needs resummation which
  makes the renormalization of the model difficult. In the paper it is
  discussed, how can one perform a consistent perturbation theory at zero as
  well as at finite temperature with the help of momentum dependent
  renormalization schemes.
\end{abstract}

\end{frontmatter}



In the O(N) model, using $1/N$ expansion we encounter non-convergent
perturbative series. The source of the problem is that at one hand the
coupling constant is proportional to the small parameter $1/N$, but, on the
other hand the number of degrees of freedom is inversely proportional to
it. As a net effect a contribution not suppressed by the small parameter can
be formed on the radiative level. This spoils the strict perturbative ordering
of loop levels, and so make the direct perturbation theory useless.

To the first level, the problem can be circumvent by introducing a new degree
of freedom with the help of Hubbard Stratonovich transformation, and then
determine the vacuum expectation value of this new field
\cite{Colemanetal}. At higher order, however, this is not enough; what helps
there is the modification of the propagation of the new degree of
freedom. Then, however, special care is needed to maintain the
renormalizability of the theory \cite{Andetal}.

In this work we use the method developed in a series of papers to perform a
renormalized 2PI resummation \cite{AJ1, AJ2}. The idea is that we split the
bare kernel into two parts, one is used as a kernel of the unperturbed theory,
the other as a counterterm. Since by construction we do not change the bare
theory, we keep the physics untouched. In \cite{AJ1} it is studied, how
consistent is the perturbation theory with a generic momentum dependent
counterterm. Shortly, the result is that if the unperturbed kernel can be
power expanded around asymptotically large momenta, then we obtain a
consistent renormalization procedure.

We start with the definition of the model: its Lagrangian reads
\begin{equation}
  {\cal L} = \frac12(\d_\mu\bar\Phi_i)(\d^\mu\bar\Phi_i)
  -\frac{\bar m^2}2\bar\Phi_i\bar\Phi_i -\frac{\bar \lambda}{24 N}
  \left(\bar\Phi_i\bar\Phi_i\right)^2.
\end{equation}
We perform a Hubbard-Stratonovich transformation to introduce a new degree of
freedom. For later convenience we perform an incomplete transformation; in
imaginary time the resulting Lagrangian reads
\begin{equation}
  {\cal L}_E =  \frac12(\d_\mu\bar\Phi_i)(\d_\mu\bar\Phi_i) + \frac{\bar
    m^2}2\bar\Phi_i\bar\Phi_i + \frac12\bar\chi^2 + \frac{i \bar g}
  {2\sqrt{N}} \bar\chi \bar\Phi_i\bar\Phi_i + \frac{\delta \lambda}{24 N}
  \left(\bar\Phi_i\bar\Phi_i\right)^2,
\end{equation}
where $\bar\lambda=3\bar g^2+\delta\lambda$. The $\delta\lambda$ term seemingly
reintroduces the bad $1/N$ behavior, however we will allow its value to be at
most ${\cal O} (1/N)$.

In the renormalization procedure must keep the bare Lagrangian intact. But we
can redefine fields and render couplings to the renormalized and counterterm
parts arbitrarily. For the field redefinition we use $\bar\Phi= Z^{1/2} \Phi$
and $\bar\chi = Z_\chi^{1/2}\chi_0 + iZ_\chi^{-1/2}\sqrt{N}\,q$, where
$Z=1+\delta Z$, $Z+\chi=1+\delta Z_\chi$, and $q$ is a c-number: it is
necessary to cancel divergences proportional to the $\chi_0$ field itself. We
split the couplings as $Z \bar m^2 = m^2+\delta m^2$ and $ZZ_\chi^{1/2}\bar g
= g+ \delta g$. To achieve resummation we split the quadratic $\chi_0$ part in
a momentum dependent way:
\begin{equation}
  \label{rest}
  \frac12 Z_\chi |\chi_0(p)|^2 = \frac12 \chi_0^*(p) H(p) \chi_0(p) + \frac12
  \chi_0^*(p) \delta H(p) \chi_0(p)
\end{equation}
where the first term is considered to be part of the free Lagrangian, the
second one is a momentum dependent counterterm. Finally we arrive at the form
\begin{eqnarray}
    {\cal L}_E && =  \frac12 \Phi_i \,(-\d^2+ m^2)\,\Phi_i + \frac12
  \chi_0\, H(i\d)\, \chi_0 + \frac{ig}{2\sqrt{N}} \chi_0\Phi_i\Phi_i +
  \sqrt{N}\,i q\chi_0  + \nn &&+  \frac12 \Phi_i \,(-\delta Z \d^2 + \delta
  m^2)\,\Phi_i + \frac12 \chi_0\,\delta H(i\d)\, \chi_0 + \frac{i\delta g}
  {2\sqrt{N}} \chi_0\Phi_i\Phi_i + \frac{\delta \lambda}{24N}
  (\Phi_i\Phi_i)^2.
\end{eqnarray}
We assume that we are in the symmetry broken phase, where $m^2<0$.

As next we introduce nontrivial background for both $\Phi$ and $\chi$. by
rotating the coordinate system in the internal space we can achieve that
\begin{equation}
    \chi_0 = -i\sqrt{N}\,X + \chi,\qquad \Phi_N = \sqrt{N} \Phi +\rh,
  \qquad \Phi_i=\ph_i\; (i=1\dots N-1).
\end{equation}

When we perform a perturbative analysis with the above Lagrangian, we find
that the $\chi$ self-energy gets ${\cal O}(N^0)$ correction. In formula we
obtain expansion
\begin{equation}
  \label{sigmachi}
  \Sigma_{\chi\chi}(p,{\cal E})= -\frac{g^2}2 I(p,{\cal E}) -\delta H_0(p),
\end{equation}
where $I$ is the finite temperature bubble diagram
\begin{equation}
  I(p,{\cal E}) = \int_q G_\pi(p-q) G_\pi(q),\qquad G_\pi(p) =
  \frac1{p^2+m_\pi^2},
\end{equation}
where $m_\pi^2 = m^2+g^2X^2$, and the integral is understood as $T
\sum_{n=-\infty}^\infty \pint3{\bf p}$. The reason for the appearance of
tree-level order correction is that although the coupling constant yields an
$1/N$ factor, the number of participants in the loop (the $\ph_i$ fields) have
a number of $N$, and the two factors cancel each other. A technically
radiative, but in reality tree-level correction is a disaster for the
perturbation theory, since this term can appear in any number as a subdiagram
in any larger diagrams, and so the number of diagrams is infinite at each
order. The common wisdom is that we have to perform a resummation in this
case.

Formula \eqref{sigmachi} suggests a solution for this problem in the present,
strictly perturbative framework. If we choose a specific scheme where
\begin{equation}
   \delta H_0(p)=-\frac{g^2}2 I(p,{\cal E}),
\end{equation}
then the $\chi$ self-energy will be zero. A concrete calculation shows that
$\delta H_0(p)$ asymptotically is $\sim \ln p$. According to \eqref{rest} the
$\chi$-kernel $H(p)$ has the same momentum dependence as the counterterm, and
so the consistency relation of \cite{AJ1} fulfills. The divergent part of
$\delta H_0(p)$ determines, through \eqref{rest}, the divergent part of
$\delta Z_\chi$. The finite parts must be determined using the renormalization
conditions, as usual.

One can also compute the free energy at the leading, ${\cal O}(N^0)$ order,
and determine the necessary counterterms, $q_0$ and the free energy zero-point
renormalization $\delta f_0$ (for details cf. \cite{AJ2}). The leading order
counterterms then read
\begin{eqnarray}
  \label{cntt}
  \delta Z_{\chi,0} && = -\frac{g^2}{32\pi^2} \ln\frac{e\Lambda^2}{m_\rh^2},
  \qquad q_0= \frac{g}{32\pi^2}\left[-\Lambda^2 + m^2
    \ln\frac{\Lambda^2}{m_\pi^2} + gX_{\mathrm{min}}
    \ln\frac{m_\rh^2}{em_\pi^2} \right],\nn
  \delta f_0&&= -\frac1{32\pi^2}\left[ m^2\Lambda^2 -\frac{m^4}2
    \ln\frac{\Lambda^2}{m_\pi^2} + \frac{g^2
      X_{\mathrm{min}}^2}2\ln\frac{m_\rh^2}{em_\pi^2} -\frac{m_\pi^4}4\right],
\end{eqnarray}
where $X_{\mathrm{min}}=-m^2/g$ is the tree level minimum of the potential,
$m_\rh^2=-2m^2$, and $m_\pi^2$ must be taken at $X_{\mathrm{min}}$. We can see
that all the counterterm are local and independent of the temperature.

After this procedure the $\chi$ self-energy is zero at the ${\cal O}(N^0)$
order, and so the normal perturbative behavior is restored. Two peculiar
remnants of the resummation problems, however, still survive. One is that the
number of $\ph_i$ fields may still lift certain contributions to higher level
of perturbation theory; but these never reach the tree level, and so there are
just a finite number of diagrams affected in this way, which does not spoil
the perturbative technique. The other peculiarity is that, as a consequence of
the nontrivial $\chi$ propagator, the divergent structures of the counterterms
may strongly deviate from the forms we are used to in the normal perturbative
cases.

In the next-to-leading order we obtain the following self-energy for the
$\ph_i$ fields:
\begin{equation}
  \label{piself}
  \hspace*{-1em} 
  \Sigma_{\pi}(p)= \frac{g^2}N\! \int_q\! G_\pi(p-q) G_{\chi\chi}(q) +\delta
  Z_1 p^2 + \delta m_1^2 +\delta g_1 X + \frac{\delta\lambda_1} 6\Phi^2 +
  \frac{\delta\lambda_1} 6\! \int_q\! G_\pi(q),
\end{equation}
where $G_{\chi\chi}(p) = (p^2+m_\pi^2)/[\,H(p)(p^2+m_\pi^2) +
g^2\Phi^2]$. Requiring finiteness we may determine the infinite parts of the
counterterms in this formula. We obtain (cf. \cite{AJ2}):
\begin{eqnarray}
  && \delta Z_1=0,\quad \delta \lambda_1=0,\quad  \delta g_1 = \frac{2g}N
  \ln\ln\frac{L^2}{\Lambda^2},\nn
  && \delta m^2_{1,div} = \frac{\Lambda^2}{N \ln L/\Lambda}\left[1 -
    \frac1{2\ln L/\Lambda} + \frac1{2\ln^2L/\Lambda}\right] + \frac{2m^2}N
  \ln\ln \frac{L^2}{\Lambda^2},
\end{eqnarray}
where $L= e^{\frac{16\pi^2}{g^2}}m_\rh$ is the UV Landau pole
position. Although these expressions are very different from the usual ones,
these are local and independent of the temperature in this order, too.

\begin{figure}[htbp]
  \centering
  \begin{minipage}[b]{4cm}
    \includegraphics[width=4cm]{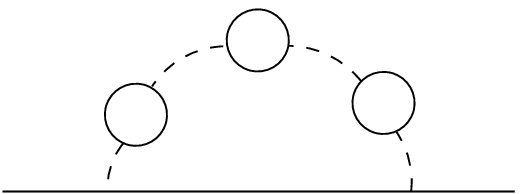}\\[0.7em]
    \centerline{a.)}
  \end{minipage}
  \hspace*{3em}
  \begin{minipage}[b]{4cm}
    \includegraphics[width=4cm]{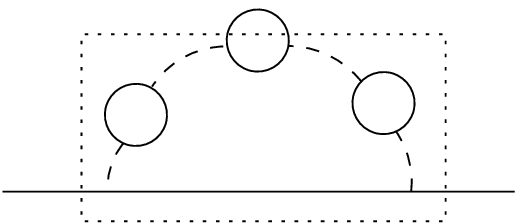}\\
    \centerline{b.)}
  \end{minipage}
  \hspace*{3em}
  \begin{minipage}[b]{2cm}
    \includegraphics[height=1.6cm]{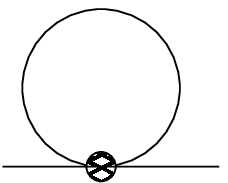}\\[0.3em]
    \centerline{c.)}
  \end{minipage}
  \caption{The pion self-energy as expanded in the language of the
    non-resummed $\chi-\chi$ propagator (a.). Its potentially divergent
    subdiagram boxed by dotted line (b.) The diagram necessary to cancel this
    subdivergence (c.)}
  \label{fig:piself}
\end{figure}

The last term in \eqref{piself} is formally a second-order contribution (one
loop with a counterterm coupling), but again the number of degrees of freedom
in the loop is $N$, which lifts up this contribution to the first order
level. Direct analysis of this order using the original (non-resummed) degrees
of freedom reveals the role of this term. The first term in \eqref{piself},
namely, represents an infinite number of diagrams, depicted on
Fig.~\ref{fig:piself}/a.
Any of these diagrams contains divergent subdiagrams, as shown by
Fig.~\ref{fig:piself}/b. The necessary counterterm is
Fig.~\ref{fig:piself}/c in all cases, which exactly corresponds to the last
term of \eqref{piself}.

As a conclusion we may say that the use of momentum dependent perturbative
schemes provide a compact way of handling resummations, with the benefit that
it also provides a tool for renormalization. In case of O(N) model with $1/N$
expansion resummation is needed only at the leading order level, and it
requires to determine the background field expectation values as well as the
correct propagator for the $\chi$ field. At higher orders usual perturbative
techniques can be used, the only remnant of a previous resummation are the
appearance of certain diagrams at lower order level and the unusual divergence
structure of the counterterms.

Acknowledgment: this work is supported by the Humboldt foundations and the
Hungarian Scientific Research Fund under contracts K-68108.



\end{document}